\documentclass[letterpaper, 10 pt]{cssconf} 

\def \a {\alpha}
\def \b {\beta}

\usepackage{amssymb}
\usepackage{amsmath}
\usepackage{graphicx,color}
\usepackage{enumerate}
\usepackage{lipsum} 
\usepackage[capitalise]{cleveref}
\usepackage{tikz}
\usetikzlibrary{patterns}
\usepackage{accents}

\usepackage[ruled]{algorithm2e}

\newtheorem{assumption}{Assumption}

\newtheorem{proposition}{Proposition}
\newtheorem{remark}{Remark}
\newtheorem{theorem}{Theorem}

\newtheorem{lemma}{Lemma}

\DeclareMathOperator{\Equaldef}{\overset{def}{=}}

\title{\LARGE \bf Improved Convergence Rate for a Distributed Two-Time-Scale Gradient Method under Random Quantization}
\author{
Marcos M. Vasconcelos,\quad Thinh T. Doan\quad and\quad Urbashi Mitra
\thanks{M. M. Vasconcelos is with the Commonwealth Cyber Initiative and the Bradley Department of Electrical and Computer Engineering, Virginia Tech. T. T. Doan is with the Bradley Department of Electrical and Computer Engineering, Virginia Tech. U. Mitra is with the Ming Hsieh Department of Electrical Engineering, University of Southern California. E-mails: \texttt{\{marcosv,thinhdoan\}@vt.edu}, \texttt{ubli@usc.edu}.}}

\begin{document}

\maketitle

\begin{abstract}
We study the so-called distributed two-time-scale gradient method for solving convex optimization problems over a network of agents when the communication bandwidth between the nodes is limited, and so information that is exchanged between the nodes must be quantized. Our main contribution is to provide a novel analysis, resulting to an improved convergence rate of this method as compared to the existing works. In particular, we show that the method converges at a rate $\mathcal{O}(\log^2(k)/\sqrt{k})$ to the optimal solution, when the underlying objective function is strongly convex and smooth. The key technique in our analysis is to consider a Lyapunov function that simultaneously captures the coupling of the consensus and optimality errors generated by the method.     
\end{abstract}

\section{Introduction}

Next generation cyber-physical systems will seamlessly incorporate machine learning capabilities to enable adaptability and resiliency to guarantee performance in non-stationary environments. In decentralized cyber-physical systems, the application of real-time, collaborative machine learning methods require a collection of agents to \textit{train} a global model efficiently based on local processing and information exchange over a network \cite{Nedic:2020}. Such problems can be formulated using the framework of distributed optimization, which has a long and rich history, see for example \cite{Yang:2019} and references therein. 

One popular class of distributed optimization algorithms consists of each agent performing gradient descent on its local function, obtaining a local estimate of the optimal solution, which is then shared with its neighbors. Such approach essentially \textit{interleaves} average-consensus and gradient descent. In contrast to quantized-consensus which has been extensively studied and is now well understood,  \textit{distributed optimization under quantization} has received significantly less attention. Since quantization is a necessary component in any application involving digital communication among agents, it is essential to design algorithms that take quantization into account and quantify the impact of quantization on the performance of such algorithms. As \cite{Nedic:2020} points out that quantization is a big bottleneck in the design and analysis of distributed optimization algorithms. Our work seeks to address this fundamental issue.

In \cite{Li:2017}, a quantized subgradient method is considered, where the quantization scheme is non-adaptive, i.e., the quantization bins do not change with time. In this case, convergence to a ball centered at the optimal solution is shown. While \cite{ReisizadehMHP2019} establishes convergence to the optimal solution with adaptive quantization, a drawback of \cite{ReisizadehMHP2019} is the need for periodic communication of the quantization intervals, which violates  hard constraints on the fixed number of bits allowed over each link, since these intervals are represented by real numbers. Such requirement is also considered in the recent work in both centralized (there exists a centralized coordinator) and distributed communication framerworks; see for example \cite{pmlr-v119-taheri20a} and the references therein. Finally, the work in \cite{9157925} provides an adaptive quantization framework to achieve an optimal rate of distributed subgradient methods under quantization, however, it requires  a (strict) condition on the capacity of the communication bandwidth; see Assumption $2$ in \cite{9157925}.   

These challenges can be overcome by two-time scale algorithms from stochastic approximation \cite{Doan:2018a,CM20}.  In particular, \cite{Doan:2018a} established convergence and a convergence rate of a distributed optimization algorithm under strict restrictions on the number of bits that are transmitted \textit{at all times}. However, that algorithm requires the use of an operator that projects the iterates onto compact sets, which are then quantized. This projection operator creates a \textit{projection error} in addition to the \textit{quantization error}. The interplay between these two quantities introduces significant complexity to the analysis and limits the achieved convergence rate. For example, the convergence rate achieved in \cite{Doan:2018a} is $\mathcal{O}(1/\sqrt[3]{k})$.





The main contribution herein is to provide a novel analysis, resulting to an improved convergence rate of the distributed two-time-scale gradient method under random quantization. In our method, the quantization bins increase with time in a deterministic way, while keeping the number of bits in communication constant at all time. This avoids the strong requirement of exchanging real numbers per communication time in the existing literature. By introducing a Lyapunov function that simultaneously captures the coupling between the consensus and optimality errors, we show that our method converges at a rate $\mathcal{O}(1/\sqrt{k})$ to the optimal solution, which is better than the one in \cite{Doan:2018a} by a factor of $1/\sqrt[6]{k}$.


\subsection{Notation}
We adopt the following notation: scalar and vector valued random variables, and corresponding realizations are denoted by lowercase letters, such as $x$. Matrices and some constants are denoted by uppercase letters, such as $X$. The all-one vector is denoted by $\mathbf{1}$. The expectation of a random variable $x$ is denoted by $\mathbf{E}[x]$. The Euclidean norm of a vector $x$ is denoted by $\|x\|$, and the Frobenius norm of a matrix $X$ is denoted by $\|X\|_F$.


\section{Problem formulation}

Consider a distributed optimization problem:
\begin{equation}\label{eq:main}
    \min_{x \in \mathbb{R}^d} f(x) \Equaldef \sum_{i=1}^nf_i(x),
\end{equation}
where each function $f_i:\mathbb{R}^d \rightarrow \mathbb{R}$ is convex, $i \in \{1,\cdots,n\}$. Each function $f_i$ is only available to agent $i$ and satisfies the following conditions. 
\begin{assumption}\label{assump:fi}
The function $f:\mathbb{R}^{d}\rightarrow\mathbb{R}$ has Lipschitz continuous gradient with constant $L$, i.e., 
\begin{equation}
\|\nabla f(x) - \nabla f(y)\| \leq L\|x-y\|,\quad \forall x,y\in\mathbb{R}^{d}.\label{assump:fi:Lipschitz}   
\end{equation}
In addition, $f_{i}$ is strongly convex with constant $\mu$, i.e., 
\begin{equation}
\frac{\mu}{2}\|x-y\|^2 \leq f(x) - f(y) - \nabla f(y)^T(x-y),\; x,y\in\mathbb{R}^{d}.\label{assump:fi:sc}   
\end{equation}
\end{assumption}
\vspace{5pt}
We also assume that the gradient of $f_i$ is bounded. 
\begin{assumption}\label{assump:bounded_grad}
There exists a constant $C>0$ such that
\begin{equation}
\|\nabla f_{i}(x)\| \leq C,\quad  x\in\mathbb{R}^{d},\; \forall i\in\{1,\cdots,n\}. \label{assump:bounded_grad:C}   
\end{equation}
\end{assumption}
\vspace{5pt}
\begin{remark}
We note that the boundedness condition of $\nabla f_{i}$ can be guaranteed through a projection step of the variable $x$ to a predefined compact set, as considered in \cite{Doan:2018a,CM20}.  For simplicity, we will consider only unconstrained problems under this assumption in this paper. However, the proposed methods extend to constrained problems as well as the relaxation of Assumption \ref{assump:bounded_grad}.
\end{remark}

Agents are connected by an undirected and static connected graph $\mathcal{G}=(\mathcal{V},\mathcal{E})$, where $\mathcal{V}=\{1,\cdots,n\}$ and ${\mathcal{E}\subseteq \mathcal{V}\times \mathcal{V}}$ are the sets of vertices and edges, respectively. Each communication link $e\in \mathcal{E}$ can support $b$ bits per dimension per transmission. Therefore, when a vector $x_i \in \mathbb{R}^d$ is transmitted from agent $i$ to one of its neighbors, it must first be quantized into a vector $q_i$ using $b\times d$ bits. We do not assume any form of analog communication among agents. The goal of the agents is to cooperatively solve the problem in  \cref{eq:main} by exchanging quantized messages over communication links of limited  bandwidth.

\subsection{Random quantization}

Here we describe the quantization algorithm employed in our framework, where each component of the local solution estimate $x_i \in \mathbb{R}^d$ is quantized using $b$ bits. The following discussion specifies the quantizer for a real number representing a single component, and the same operation is repeated for all the $d$ dimensions of $x_i$.


Let $x\in [\ell,u]$. We partition the interval $[\ell,u]$ into $B$ equal length bins whose endpoints are denoted by $\tau_m$, $m\in\{1,\cdots, B+1\}$, such that $\tau_1=\ell$, and $\tau_{B+1}=u$. Define the length of each bin by $\Delta$, such that:
 \begin{equation}
     \Delta\Equaldef \frac{u-l}{B}.
 \end{equation}
 We use $\{\tau_m\}_{m=1}^{B+1}$ as the representation symbols for the quantizer. As such, each $\tau_m$ is mapped into a codeword of $b$ bits. Hence, for a given number of bits $b$, the number of bins is given by $B=2^b-1$, which implies that:
 \begin{equation}\label{eq:interval}
     \Delta = \frac{u-l}{2^b-1}.
 \end{equation}
 
 Let $x\in[\tau_m,\tau_{m+1})$, then we choose either $\tau_m$ or $\tau_{m+1}$ as a representation point for $x$ based on the following stochastic rule, $\mathcal{Q}$, defined as:
 \begin{equation}
     \mathcal{Q}(x) \Equaldef \begin{cases}
     \tau_m &  \ \ \text{with prob.}\  1-(x-\tau_m)/\Delta\\
     \tau_{m+1} &  \ \ \text{with prob.}\  (x-\tau_m)/\Delta,
     \end{cases}
 \end{equation}
 where $\Delta$ is given by \cref{eq:interval}.


The randomized quantizer $\mathcal{Q}$ described above satisfies three important properties:
\begin{equation}\label{eq:unbiased}
  \mathbf{E}[\mathcal{Q}(x) \mid x]= x,
\end{equation}
\begin{equation}
\mathbf{E}\big[\big(\mathcal{Q}_b(x)-x\big)^2 \mid x\big]\leq \frac{\Delta^2}{4},
\end{equation}
and
\begin{equation}\label{eq:bounded_support}
   \mathbf{P} \Big( \big|\mathcal{Q}(x)-x\big| \leq \Delta \Big) =1.
\end{equation}

\subsection{Distributed two-time scale  gradient method under random quantization}

We are interested in exploiting distributed two-time-scale gradient methods, formally stated in Algorithm \ref{alg:QDSG}, for solving  problem in \cref{eq:main} under random quantization. This algorithm has a simple interpretation given as follows.

Each agent $i\in \mathcal{V}$ keeps a local estimate of the solution to \cref{eq:main}. At time step $k$, the $i$-th agent's local estimate is denoted by $x^i_k$. Each agent has access to a local subgradient of its local function. At time step $k$, the $i$-th agent's local gradient at $x^i_k$ is denoted by $g^i_k$, i.e., 
\begin{equation}
    g_i^k \Equaldef  \nabla f_i(x^k_i).
\end{equation} 
At time $k$ the $i$-th agent received the quantized versions of the states from every agent $j\in\mathcal{N}_i$. Let
\begin{equation}\label{eq:algorithm}
    q^j_k \Equaldef \mathcal{Q}\big(x^j_k\big)
\end{equation}
denote the quantized version of the local estimate of the $j$-th agent at time $k$. Each agent $i$ then iteratively updates its variable as 
\begin{equation} \label{eq:main_algorithm}
    x^i_{k+1} \Equaldef 
    \big(1-\beta_k \big)x^i_k+\beta_k\sum_{j\in \mathcal{N}_i}a_{ij}q^j_k  -\alpha_kg_k^i, \ \ i\in \mathcal{V},
\end{equation}
where $\{\a_k\}_{k=0}^\infty$ and $\{\b_k\}_{k=0}^\infty$ are sequences of diminishing step-sizes and $a_{ij}$ are averaging coefficients corresponding to the entries of a doubly stochastic matrix $A$. The iterates generated by \cref{eq:main_algorithm} will be the centerpiece of this paper, for which we will establish the convergence rate under appropriate assumptions on the structure of the problem to be specified in the sequel.

\begin{algorithm}[t]
\caption{Distributed Two-Time-Scale Gradient Methods Under Random Quantization}
\begin{enumerate}
\item \textbf{Initialize}: Each node $i$ initializes $x^i_0 = 0$, and two sequences of stepsizes $\{\alpha_k,\beta_k\}_{k\in\mathbb{N}}$.\vspace{0.2cm}
\item \textbf{Iteration}:  For $k=0,1,\cdots,$ node $i\in\mathcal{V}$ implements:
\begin{itemize}[leftmargin = 4mm]
\item[a.] Compute random quantization $q^{i}_{k} = \mathcal{Q}(x^{i}_{k})$
\item[b.] Send $q^{i}_{k}$ to node $j\in\mathcal{N}_i$
\item[c.] Receive $q^{j}_{k}$ from node $j\in\mathcal{N}_i$ and update
\begin{align*}
x^i_{k+1} = 
(1-\beta_k)x^{i}_{k} + \beta_{k}\sum_{j\in\mathcal{N}_i} a_{ij}q^{j}_{k} - \alpha_{k} g^i_{k}.
\end{align*}
\item[d.] Update the output $z^{i}_{k+1}$
\begin{align*}
z^i_{k+1} = \frac{\sum_{t=0}^{k}(t+1) x^i_{t}}{\sum_{t=0}^{k}(t+1)}\cdot
\end{align*}
\end{itemize}
\end{enumerate}
\label{alg:QDSG}
\end{algorithm}

\subsection{Quantization with increasing bin size}\label{sec:increasing_quantizer}

What enables the analysis of the algorithm in \cref{eq:main_algorithm} is the following proposition, which guarantees that there exists a randomized quantization scheme with increasing bin sizes that has a maximal error that grows logarithmically with $k$. The purpose of Proposition 1 is to allow each agent to know at any given $k$ which interval to quantize and decode. By using the construction given in the proof, we only need to transmit $b\times d$ bits every iteration but not a real number as done in \cite{ReisizadehMHP2019,pmlr-v119-taheri20a}. 

\begin{proposition}\label{prop:quantization}
Let $\{\a_k\}$ be a sequence of diminishing step-sizes in algorithm \cref{eq:main_algorithm}. There exists a randomized quantization scheme such that: 
\begin{equation}
    \|x^i_k -q^i_k\| \leq \Delta_k \Equaldef \frac{2C}{2^b-1}\sum_{t=0}^{k-1}\a_t
\end{equation}
\end{proposition}
\vspace{5pt}
\begin{proof}
Let $x^0_i=0$, then since $|g_i^0|\leq C$ for all $i$, we have:
\begin{equation}
    |x_1^i|\leq \a_0C.
\end{equation}
At $k=1$, let agent $i$ quantize $x^i_1 \in [-\a_0C,\a_0C]$ to obtain $q_1^i$. Then for all $i$
\begin{equation}\label{eq:prop_quant}
    \mathbf{E}[q^1_i \mid x^1_i]=x^1_i \ \ \text{and} \ \ |x^1_i - q^1_i| \leq \Delta_1,
\end{equation}
where 
\begin{equation}
    \Delta_1 = \frac{2\a_1C}{2^b-1}.
\end{equation}
Then, at $k=1$, we compute $x_i^{2}$ using \cref{eq:main_algorithm} as:
\begin{equation}
        x^i_{2} \Equaldef 
    \big(1-\beta_1 \big)x^i_1+\beta_1\sum_{j\in \mathcal{N}_i}a_{ij}q^j_1  -\alpha_1g_1^i
\end{equation}
Thus, since $\{ |x^i_1|,|q^i_1|\}_{i=1}^n \leq \a_0C$,  we have
\begin{IEEEeqnarray}{rCl}
    |x^i_2| & \stackrel{(a)}{\leq} & (1-\beta_1)|x^i_1| + \beta_1 \sum_{j\in \mathcal{N}_i}a_{ij}|q^j_1| + \alpha_1|g^i_1| \\
    & \stackrel{(b)}{\leq} & C(\a_0 + \a_1), 
\end{IEEEeqnarray}
where $(a)$ follows from the triangle inequality, and $(b)$ follows from \cref{eq:prop_quant}.
Therefore, quantizing $x^2_i \in [-(\a_0+\a_1)C,(\a_0+\a_1)C]$, we have
\begin{equation}
    |x^i_2 - q^i_2| \leq \Delta_2 = \frac{2C}{2^b-1}(\a_0+\a_1).
\end{equation}

By induction, repeating the steps above for the update and quantization, we obtain:
\begin{equation}
    |x^i_k - q^i_k|\leq \Delta_k = \frac{2C}{2^b-1}\sum_{t=0}^{k-1}\a_t.
\end{equation}
\end{proof}

\begin{remark}
For strongly convex functions, we will later choose $\a_k \propto 1/(k+1)$, thus the quantization error will satisfy:
\begin{equation}
    \Delta_k \approx \frac{2C}{2^b-1}\ln (k)
\end{equation}
\end{remark}
\vspace{5pt}
Proposition 1 is proven for scalars, but the same analysis can be carried out for vectors, for which the error bound is simply multiplied by the appropriate dimension $d$, i.e.,
\begin{equation}
    \Delta_k = \frac{2dC}{2^b-1}\sum_{t=0}^{k-1}\a_t.
\end{equation}

The main consequence of \cref{prop:quantization} is that we do not need to use an additional projection step as in \cite{Doan:2018b}. Unlike \cite{Doan:2018b}, the quantization error now depends on time step $k$ instead of being constant. This error scales with $\ln(k)$, and fortunately, will not impact the asymptotic convergence rate.

\subsection{Preliminary definitions and alternative representations}
Throughout the analysis of the algorithm, one quantity will be paramount: the  \textit{quantization error} process. For the $i$-th agent let the quantization error be defined as:
\begin{equation}
    e^i_k \Equaldef x^i_k - q^i_k.
\end{equation}

To facilitate the analysis, we will use $X,G(X)\in \mathbb{R}^{n\times d}$ to denote the following matrices:
\begin{equation}
    X = \begin{bmatrix} 
    -  x^T_1  - \\ 
    \cdots \\
    -  x^T_n  -
    \end{bmatrix} \ \ \ \text{and} \ \ \ G(X) = \begin{bmatrix} 
    -  g^T_1(x_1)  - \\ 
    \cdots \\
    -  g^T_n(x_n)  -
    \end{bmatrix}.
\end{equation}

Due to the random quantization operator, the sample-paths $x^i_k$ originating from the recursion in \cref{eq:main_algorithm} are stochastic processes. Let $\mathcal{F}_k$ be the filtration \cite{Cinlar:2011} containing all of the history generated by \cref{eq:main_algorithm} up to time $k$:
\begin{equation}
    \mathcal{F}_k \Equaldef \big\{X_0,Q_0,X_1,Q_1,\cdots, X_k,Q_k\big\}.
\end{equation}
The conditional expectation operator with respect to the filtration $\mathcal{F}_k$ is denoted by $\mathbf{E}_{\mathcal{F}_k}$.

Let the average of $x^i_k$'s be denoted by $\bar{x}_k $, such that
\begin{equation}
    \bar{x}_k \Equaldef \frac{1}{n}\sum_{i=1}^nx^i_k = X_k^T\mathbf{1}. 
\end{equation}
Finally, define the following stochastic processes:
\begin{equation}
    r_k \Equaldef \|\bar{x}_k-x^*\|
\end{equation}
and
\begin{equation}\label{eq:Y_def}
    Y_k \Equaldef X_k - \mathbf{1}\bar{x}_k^T = WX_k,
\end{equation}
where $W = I - \frac{1}{n}\mathbf{1}\mathbf{1}^T$.

In our analysis of algorithm \cref{eq:algorithm} we will also make use of the two following alternative representations:
\begin{equation}\label{eq:alt_rep_1}
    \bar{x}_{k+1}= (1-\b_k)\bar{x}_k + \b_k\bar{q}_k - \a_k\bar{g}_k,
\end{equation}
and 
\begin{equation}\label{eq:alt_rep_2}
    X_{k+1} = (1-\b_k)X_k + \b_kAQ_k - \a_kG_k.
\end{equation}


\section{Main result}
Our main contribution is to show that under the strict quantized communication constraints, we can improve upon the previously achievable convergence rates.




\begin{theorem}\label{thm:main}
Suppose that Assumptions \ref{assump:fi} and \ref{assump:bounded_grad} hold. Let $\{x_{i}^{k}\}$ be generated by Algorithm \ref{alg:QDSG}, and $\{\alpha_{k},\beta_{k}\}$ satisfy
\begin{equation}
\a_k = \frac{4/\mu}{k+1},\quad 
\b_k = \frac{4/(1-\sigma_2)}{(k+1)^{3/4}},\label{thm:main:stepsizes}
\end{equation}
where $\sigma_2$ is the second largest eigenvalue of the adopted averaging matrix $A$. Then the output $z_{i}^{k}$ of Algorithm \ref{alg:QDSG} at each agent $i$ satisfies
\begin{multline}\label{thm:main:rate}
\mathbf{E}\left[f(z_k^i)\right] - f^*  \leq \\
    \mathcal{O}\left(\frac{\mathbf{E}\|x_{0} - x^*\|^2}{k^2} 
    +  \frac{n^2(\ln k)^2}{(1-\sigma_2)^2k^{3/4}} + \frac{(\ln k)^2}{(1-\sigma_{2})\sqrt{k}}\right).
\end{multline}
\end{theorem}
\begin{remark}
An exact expression for the rate in \eqref{thm:main:rate} is given in \eqref{eq:rate}. Here we provide key observations on the result presented in Theorem \ref{thm:main}.

We note that the convergence rate of Algorithm \ref{alg:QDSG} in \eqref{thm:main:rate} is better than those in \cite{ReisizadehMHP2019,Doan:2018a} (where we ignore the log factor). In particular, a convergence rate $\mathcal{O}(1/k^{(1-\gamma)/2})$ for some $\gamma\in(0,1)$ is shown in \cite{ReisizadehMHP2019}. However, this parameter $\delta$ impacts the speed of the algorithm as mentioned by the authors (see the discussion after Theorem $1$ in \cite{ReisizadehMHP2019}). Specifically, as $\delta$ goes to zero, they achieve a rate close to $1/\sqrt{k}$, at the cost of spending more iterations due to some constants in their rates getting larger. It is also worth to recall that this work requires the communication of real numbers between agents at every iteration. On the other hand, a convergence rate $\mathcal{O}(1/k^{1/3})$ is provided in \cite{Doan:2018a}. 

We note that for unquantized (infinite bandwith) problems with strongly convex objectives, distributed consensus-based gradient methods achieve $\mathcal{O}(1/k)$ and linear convergence rates for non-smooth and smooth functions, respectively, see for example   \cite{nedic2018network}. Thus, quantization does slow down the rate of convergence of this method.
\end{remark}

\section{Analysis}

In this section, we state and prove two technical Lemmas that will be used to prove Theorem 1.

\begin{lemma}\label{lem:bound_Y} Let the sequence $\{x^i_k\}$ be generated by \cref{eq:main_algorithm}, $i\in \mathcal{V}$. Let $\{\a_k,\b_k\}$ be two sequences of non-negative and non-increasing step sizes. Let $(1-\sigma_2)$ be the spectral gap of the network connectivity. If $f$ is $L$-Lipschitz continuous, then
\begin{multline}
    \mathbf{E}_{\mathcal{F}_k} \|Y_{k+1}\|_F^2 \leq \big(1-(1-\sigma_2)\b_k\big)\|Y_k\|_F^2 \\
    + \big(1+(1-\sigma_2)\b_0\big)\b^2_k\sigma_2^2n\Delta_k^2 \\ +  \Big(\frac{(1-\sigma_2)\beta_0+1}{1-\sigma_2}\Big)L^2\frac{\a_k^2}{\b_k}.
\end{multline}
\end{lemma}
\vspace{5pt}
\begin{proof}
Beginning with the definition of $Y_k$, we obtain the following sequence of identities:
\begin{IEEEeqnarray*}{rCl}
\|Y_{k+1}\|_F^2 & = &  \|WX_{k+1}\|_F^2 \\
& \stackrel{(a)}{=} & \| W\big((1-\b_k)X_k + \b_kAQ_k - \a_kG_k\big) \|^2_F \\
& \stackrel{(b)}{=}  & \| (1-\b_k)Y_k + \b_kAWQ_k - \a_kWG_k \|^2_F\\
& \stackrel{(c)}{=} & (1-\eta)\underbrace{\| (1-\b_k)Y_k + \b_kAWQ_k\|^2_F}_{\Equaldef (\diamondsuit)}
\\
& & \hspace{72pt}+ \Big(1+\frac{1}{\eta}\Big)\a_k^2\| WG_k\|^2_F
\end{IEEEeqnarray*}
where $(a)$ follows from \cref{eq:alt_rep_2}, $(b)$ follows from the fact that $WA=AW$, and $(c)$ follows from \cref{prop:very_useful} in Appendix A, where $\eta$ is an arbitrary positive constant that will be specified later in the proof. Let us analyse the term $(\diamondsuit)$, starting with the following identity:
\begin{IEEEeqnarray*}{rCl}
    (\diamondsuit) & = & \|(1-\b_k)Y_k + \b_kAW(X_k+E_k)) \|^2_F \\
    & = & \|\big((1-\b_k)I + \b_kA\big)Y_k \|^2_F + \b_k^2\|AWE_k \|^2_F
\\ & & \hspace{20pt} + 2 \big\langle \big((1-\b_k)I + \b_kA\big)Y_k,\b_kAWE_k  \big\rangle_F.
\end{IEEEeqnarray*}

Taking the conditional expectation with respect to the filtration $\mathcal{F}_k$, we obtain:
\begin{IEEEeqnarray*}{rCl}
    \mathbf{E}_{\mathcal{F}_k}(\diamondsuit) & \stackrel{(a)}= & \|\big(I - (I - A)\b_k\big)Y_k \|^2_F +  \b_k^2\mathbf{E}_{\mathcal{F}_k} \|AWE_k \|^2_F \\
    & \stackrel{(b)}{\leq} &\big(1-(1-\sigma_2)\b_k\big)^2\| Y_k\|^2_F + n^2\b^2_k \Delta^2_k,
\end{IEEEeqnarray*}
where (a) follows from the unbiasedness of the random quantization scheme, $(b)$ follows from
\begin{IEEEeqnarray}{rCl}
    \|AWE_k \|^2_F & \leq & \|AW\|^2_F\|E_k\|^2_F  \stackrel{(c)}{\leq}  n^2\Delta_k^2,
\end{IEEEeqnarray}
and $(c)$ follows from
\begin{equation}
    \|AW\|^2_F \leq \|W\|^2_F = n-1 \leq n.
\end{equation}
Therefore, 
\begin{multline}
    \mathbf{E}_{\mathcal{F}_k} \|Y_{k+1}\|^2 \leq (1+\eta)\Big[ \big(1-(1-\sigma_2\big)\b_k\big)^2\|Y_k\|^2_F + n^2\b^2_k \Delta^2_k\Big]  \\ + \Big(1+\frac{1}{\eta}\Big)\a_ k^2L^2
\end{multline}
Setting $\eta = (1-\sigma_2)\b_k$, we obtain
    \begin{multline}
    \mathbf{E}_{\mathcal{F}_k} \|Y_{k+1}\|^2 \leq \big(1+(1-\sigma_2)\b_k\big)\Big[ \big(1-(1-\sigma_2)\b_k\big)^2 \|Y_k\|^2_F \\ +  2\b_k^2n^2\Delta^2_k \Big]  + \Big(1+\frac{1}{(1-\sigma_2)\b_k}\Big)\a_k^2L^2.
\end{multline}

Next, consider the following sequence of inequalities: 
\begin{equation}
    \| WG_k\|^2_F \leq  \|G_k\|^2_F \leq \sum_{i=1}^nL_i^2 \leq L^2. 
\end{equation}
After some algebraic manipulations, we get
\begin{multline}
    \mathbf{E}_{\mathcal{F}_k} \|Y_{k+1}\|^2 \leq \big(1-(1-\sigma_2)^2\b_k\big)  \big(1-(1-\sigma_2)\b_k\big)^2 \|Y_k\|^2_F \\ +  \big(1+(1-\sigma_2)\b_k\big)  n^2\b_k^2\Delta^2_k \\ + \Big(\frac{(1-\sigma_2)\b_k+1}{(1-\sigma_2)}\Big)L^2\frac{\a_k^2}{\b_k}.
\end{multline}
Finally, using the fact that $\{\b_k\}$ is a nonincreasing sequence, we obtain the following inequality
    \begin{multline}
    \mathbf{E}_{\mathcal{F}_k} \|Y_{k+1}\|^2 \leq \big(1-(1-\sigma_2)\b_k\big) \|Y_k\|^2_F \\ +  \big(1-(1-\sigma_2)\b_0\big)n^2\b_k^2\Delta^2_k   \\ + \Big(\frac{(1-\sigma_2)\b_0+1}{(1-\sigma_2)}\Big)L^2\frac{\a_k^2}{\b_k}.
\end{multline}
\end{proof}

For the second technical result in this section, we will obtain an upper bound on the conditional expectation of $r_{k+1}$ with respect to the filtration $\mathcal{F}_k$. To do so, we assume strong convexity of the global objective function, $f$. \begin{lemma}\label{lem:bound_r}
Assume that $f$ is a $\mu$-strongly convex function with $L$-Lipschitz continuous gradient. Let $\{\a_k\}$ be a non-increasing, non-negative sequence, and
\begin{equation}
    r_{k+1} = \|\bar{x}_{k+1}-x^*\|^2.
\end{equation}
Then, the following inequality holds:
\begin{multline}
    \mathbf{E}_{\mathcal{F}_k} [r_{k+1}] \leq \Big(1-\frac{\mu}{2}\a_k \Big) r_k + \a_k^2L^2 +\b_k^2\Delta_k^2 \\
+2\a_k \big((f(x^*) - f(x_k^l)\big)  + \a_k(L+\frac{8L^2}{\mu})\|Y_k\|_F^2
\end{multline}
for all $l \in \mathcal{V}$.
\end{lemma}
\vspace{5pt}
\begin{proof}
We begin with the following identity:
\begin{IEEEeqnarray}{rCl}
   \| \bar{x}_{k+1}-x^*\|^2 &=& \|(1-\b_k)\bar{x}_k + \b_k\bar{q}_k - \a_k\bar{g}_k-x^* \|^2 \nonumber\\
   & = & \underbrace{\|\bar{x}_k -x^* -\a_k\bar{g}_k\|^2}_{\Equaldef{A}} + \underbrace{\b_k^2 \|\bar{e}_k\|^2}_{\Equaldef B} \nonumber \\
  & & \hspace{25pt} + \underbrace{2 \b_k\big( \bar{x}_k -x^* -\a_k\bar{g}_k\big)^T\bar{e}_k}_{\Equaldef C}. 
\end{IEEEeqnarray}
We proceed to analyse each of the terms $A$, $B$, and $C$ above.

First, consider
\begin{equation}
    A \Equaldef \|\bar{x}_k -x^* -\a_k\bar{g}_k \|^2
\end{equation}
Then, the following identity holds:
\begin{multline}
    A = \|\bar{x}_k - x^*\|^2 + \a_k^2\|\bar{g}_k\|^2 \\\underbrace{-2\a_k\frac{1}{N}\sum_{i=1}^N \nabla f_i(x^i_k)^T(\bar{x}_k-x^*)}_{\Equaldef A_1}.
\end{multline}
Consider the following identity
\begin{multline}
A_1 = \underbrace{-2\a_k\frac{1}{n}\sum_{i=1}^n \Big(\nabla f_i(x^i_k) -\nabla f_i(\bar{x}_k) \Big)^T(\bar{x}_k-x^*)}_{\Equaldef A_{11}} \\ \underbrace{-2\a_k  \nabla f(\bar{x}_k)^T(\bar{x}_k-x^*)}_{\Equaldef A_{12}}.
\end{multline}
Then the following sequence of inequalities hold:
\begin{IEEEeqnarray}{rCl}
A_{11} & \stackrel{(a)}{\leq} & 2\a_k\frac{1}{n}\sum_{i=1}^n \|\nabla f_i(x^i_k) -\nabla f_i(\bar{x}_k)\|\|\bar{x}_k -x^*\| \\
& \stackrel{(b)}{\leq} & 2\a_k\frac{1}{n}\sum_{i=1}^n L \|x^i_k - \bar{x}_k\|\|\bar{x}_k -x^*\|\\
& \stackrel{(c)}{\leq} & \a_k\frac{L}{n}\sum_{i=1}^n  \Big(\eta\|x^i_k - \bar{x}_k\|^2 + \frac{1}{\eta}\|\bar{x}_k -x^*\|^2 \Big)\\
& \stackrel{(d)}{=} & \a_k\eta L  \|X_k - \mathbf{1}\bar{x}_k\|_F^2 + \a_kL\frac{1}{\eta}\|\bar{x}_k -x^*\|^2,
\end{IEEEeqnarray}
where $(a)$ follows from the Cauchy-Schwarz inequality, $(b)$ follows from the Lipschitz continuity of $f_i$, $i=1,\cdots,n$, $(c)$ follows from \cref{prop:very_useful}, and $(d)$ follows from \cref{eq:Y_def}. Setting $\eta = 4L/\mu$, we obtain:
\begin{equation}
    A_{11} \leq \a_k\frac{4L^2}{\mu}\|Y_k\|^2_F + \a_k\frac{4}{\mu}\|\bar{x}_k-x^*\|^2.
\end{equation}

We proceed to bounding $A_{12}$ as follows:
\begin{IEEEeqnarray}{rCl}
A_{12} & = &  2\a_k  \nabla f(\bar{x}_k)^T(x^*-\bar{x}_k) \\
& \stackrel{(a)}{\leq} & 2\a_k\Big(f(x^*) - f(\bar{x}_k)-\frac{\mu}{2}\|x^* - \bar{x}_k\|^2) \Big) \\
& \stackrel{(b)}{=} & 2\a_k \big(f(x^*)-f(x^l_k)+f(x^l_k)-f(\bar{x}_k) \big) \nonumber \\
& & \hspace{100pt} -\a_k\mu\|x^* - \bar{x}_k\|^2 \\
& \stackrel{(c)}{\leq} & 2\a_k \big(f(x^*) - f(x_k^l)) \nonumber \\  & & + 2\a_k\Big( \nabla f(\bar{x}_k)^T (x^l_k-\bar{x}_k)   +\frac{L}{2}\|x^l_k-\bar{x}_k\|^2 \Big) \nonumber \\
& & \hspace{100pt} -\a_k\mu\|x^* - \bar{x}_k\|^2 \\
& \stackrel{(d)}{=} & 2\a_k \big(f(x^*) - f(x_k^l)) \nonumber \\  & & + 2\a_k\Big( \big(\nabla f(\bar{x}_k) - \nabla f(x^*) \big)^T (x^l_k-\bar{x}_k)    \Big) \nonumber \\
& & \hspace{30pt} +\a_kL\|x^l_k-\bar{x}_k\|^2-\a_k\mu\|x^* - \bar{x}_k\|^2 \\
& \stackrel{(e)}{\leq} & 2\a_k \big(f(x^*) - f(x_k^l)) \nonumber 
\\ & & \hspace{20pt}+ 2\a_kL \| \bar{x}_k - x^* \| \|x^l_k-\bar{x}_k\|     \nonumber \\
& & \hspace{30pt} +\a_kL\|x^l_k-\bar{x}_k\|^2-\a_k\mu\|x^* - \bar{x}_k\|^2\\
& \stackrel{(f)}{\leq} & 2\a_k \big(f(x^*) - f(x_k^l)) \nonumber 
\\ & & \hspace{20pt}+ \a_kL \Big( \eta\| \bar{x}_k - x^* \|^2 + \frac{1}{\eta} \|x^l_k-\bar{x}_k\|^2\Big)     \nonumber \\
& & \hspace{30pt} +\a_kL\|x^l_k-\bar{x}_k\|^2-\a_k\mu\|x^* - \bar{x}_k\|^2,
\end{IEEEeqnarray}
where $(a)$ follows from $f$ being $\mu$-strongly convex, $(b)$ follows from adding and subtracting $f(x^l_k)$, $(c)$  follows the $L$-Lipschitz continuity of the gradient, and $(d)$ follows from the first order optimality condition $\nabla f(x^*)=0$, $(e)$ follows from the Cauchy-Schwarz inequality, and $(f)$ follows from \cref{prop:very_useful}. Setting $\eta = 4L/\mu$, we obtain:

\begin{multline}
    A_{12} \leq 2\a_k \big(f(x^*) - f(x_k^l)) - \a_k\frac{3\mu}{4}\|x^* - \bar{x}_k\|^2 \\ + \a_k\Big( L + \frac{4L^2}{\mu}\Big)\|x^l_k-\bar{x}_k\|^2.
\end{multline}

Therefore,
\begin{multline}
    A_1 \leq 2\a_k \big((f(x^*) - f(x_k^l)\big) - \a_k\frac{\mu}{2} \|\bar{x}_k-x^*\|^2 \\ + \a_k(L+\frac{4L^2}{\mu})\|x^l_k - \bar{x}_k\|^2 + \a_k\frac{4L^2}{\mu}\|Y_k\|_F^2.
\end{multline}
We next turn our attention to $B$. From the convexity of $\|\cdot\|^2$, we obtain the following upper bound:
\begin{equation}
    B = \b^2_k \Big\|\frac{1}{n}\sum_{i=1}^ne^i_k\Big\|^2 \leq \b^2_k \frac{1}{n}\sum_{i=1}^n\|e^i_k\|^2.
\end{equation}

Finally, taking the expectation of $\|\bar{x}_{k+1}-x^*\|^2$ with respect to $\mathcal{F}_k$, the term $C$ vanishes due to the fact that the random quantizer is unbiased \cref{eq:unbiased}. Moreover, due to \cref{eq:bounded_support} the term $B$ is upper bounded by
\begin{equation}
    \mathbf{E}_{\mathcal{F}_k}[B] \leq \b_k^2\Delta_k^2.
\end{equation}

Thus,
\begin{multline}
    \mathbf{E}_{\mathcal{F}_k}\|\bar{x}_{k+1}-x^*\|^2 \leq  \|\bar{x}_k - x^*\|^2 + \a_k^2\|\bar{g}_k\|^2 +\b_k^2\Delta_k^2 \\
+2\a_k \big((f(x^*) - f(x_k^l)\big) - \a_k\frac{\mu}{2} \|\bar{x}_k-x^*\|^2 \\ + \a_k(L+\frac{4L^2}{\mu})\|x^l_k - \bar{x}_k\|^2 + \a_k\frac{4L^2}{\mu}\|Y_k\|_F^2. 
\end{multline}

Finally, since
\begin{equation}
    \Big\|\frac{1}{n}\sum_{i=1}^ng^i_k\Big\|^2 \leq \frac{1}{n} \sum_{i=1}^n \|g^i_k\|^2 \leq L^2,
\end{equation}
we have:
\begin{multline}
    \mathbf{E}_{\mathcal{F}_k} [r_{k+1}] \leq \Big(1-\frac{\mu}{2}\a_k \Big) r_k + \a_k^2L^2 +\b_k^2\Delta_k^2 \\
+2\a_k \big((f(x^*) - f(x_k^l)\big)  + \a_k(L+\frac{4L^2}{\mu})\|x^l_k - \bar{x}_k\|^2 \\+ \a_k\frac{4L^2}{\mu}\|Y_k\|_F^2. 
\end{multline}
\end{proof}

\section{Proof of Theorem 1 and convergence rate improvement}
We begin by specifying the step-size sequences 
\begin{equation}
    \a_k = \frac{4/\mu}{k+1}
\end{equation}
and
\begin{equation}
    \b_k = \frac{4/(1-\sigma_2)}{(k+1)^{3/4}}.
\end{equation}

\begin{lemma}\label{lem:lyapunov}Let $\{\eta_k\}$ be a non-increasing sequence defined as:
\begin{equation}
    \eta_k \Equaldef \eta \frac{\alpha_k}{\beta_k}.
\end{equation} 
Define the following Lyapunov function:
\begin{equation}
    \mathcal{V}_{k+1} \Equaldef  r_{k+1} + \eta_{k+1}\|Y_{k+1}\|_F^2.
\end{equation}

The following inequality holds
\begin{multline}\label{eq:lyapunov_ineq}
    \mathbf{E}\big[\mathcal{V}_{k+1} \big] \leq \Big(\frac{k-1}{k+1} \Big) \mathbf{E}\big[\mathcal{V}_k\big] \\ - \Big(
    \frac{8/\mu}{k+1}\Big) \big( f(x^l_k)-f(x^*) \big) 
    + \Gamma_k,
\end{multline}
where
\begin{multline}
    \Gamma_k = \frac{16/\mu^2}{(k+1)^ 2} + \frac{40L^ 2(L+8L^2/\mu)/\mu^ 3}{(k+1)^{3/2}} \\
    + \Big[ \frac{4}{(1-\sigma_2)(k+1)^ {3/2}} + \frac{320(L+8L^ 2/\mu)n^2}{(1-\sigma_2)^ 2(k+1)^{7/4}} \Big]\\\times\Big(\frac{Cd}{2^b-1}\Big)a^2\Big( \sum_{t=0}^{k-1}\a_k\Big)^2.
\end{multline}
\end{lemma}

\vspace{5pt}

\begin{proof}
    The proof is in Appendix B.
\end{proof}

Equipped with Lemmas 1, 2, and 3, we are now ready to prove Theorem 1. 

\vspace{5pt}

\begin{proof}\textit{(Proof of Theorem 1)}

From \cref{lem:lyapunov}, multiplying both sides of  \cref{eq:lyapunov_ineq} by $(k+1)^2$ yields:
    \begin{multline}\label{eq:lyapunov_ineq2}
    (k+1)^2\mathbf{E}\big[\mathcal{V}_{k+1} \big] \leq (k^2-1) \mathbf{E}\big[\mathcal{V}_k\big] \\ - 
   (8/\mu)(k+1) \big( f(x^l_k)-f(x^*) \big) 
    + (k+1)^2\Gamma_k,
\end{multline}
which, due to the non-negativity of $\mathcal{V}_k$, can be further relaxed to:
    \begin{multline}\label{eq:lyapunov_ineq2}
    (k+1)^2\mathbf{E}\big[\mathcal{V}_{k+1} \big] \leq k^2 \mathbf{E}\big[\mathcal{V}_k\big] \\ - 
   (8/\mu)(k+1) \big( f(x^l_k)-f(x^*) \big) 
    + (k+1)^2\Gamma_k.
\end{multline}
Summing both sides of \cref {eq:lyapunov_ineq2} from $k=1$ to $T>1$, we get:
 \begin{multline}\label{eq:lyapunov_ineq2}
    (T+1)^2\mathbf{E}\big[\mathcal{V}_{T+1} \big] \leq  \mathbf{E}\big[\mathcal{V}_1\big]  \\- 
   (8/\mu)\sum_{k=1}^T(k+1) \big( f(x^l_k)-f(x^*) \big)
    + \sum_{k=1}^T(k+1)^2\Gamma_k.
\end{multline}
We will now analyze $\sum_{k=1}^T(k+1)^2\Gamma_k$. First, notice that:
\begin{equation}
    \sum_{t=0}^{T-1}\a_t = \frac{4}{\mu}\sum_{t=0}^{T-1}\frac{1}{t+1}\stackrel{(a)}{\leq} \frac{4}{\mu}\int_{0}^{T-1}\frac{1}{t+1}dt = \frac{4}{\mu}\ln T,
\end{equation}
where $(a)$ follows from the Cauchy-Maclaurin \textit{integral test}. Then, we consider
\begin{equation}
     \sum_{k=1}^T (k+1)^{1/2} \stackrel{(b)}{\leq} \int_1^T(t+1)^{1/2}dt \stackrel{(c)}{\leq} \frac{2}{3}(T+1)^{3/2},
\end{equation}
where $(b)$ follows from the integral test, and $(c)$ is obtained by dropping the term corresponding to the lower-end of the integration interval.
Therefore, we have:
\begin{IEEEeqnarray}{rCl}
    \sum_{k=1}^T (k+1)^{1/2}\bigg(\sum_{t=0}^{k-1}\a_t\bigg)^2 & \stackrel{(d)}{\leq} &     \sum_{k=1}^T (k+1)^{1/2}\bigg(\sum_{t=0}^{T-1}\a_t\bigg)^2 \\
    & \stackrel{(e)}{\leq} & \frac{32}{3\mu^2}(\ln T)^2 (T+1)^{3/2}.
\end{IEEEeqnarray}
where $(d)$ follows from $k\leq T$, and $(e)$ follows from the integral test.

Similarly, we obtain
\begin{equation}
      \sum_{k=1}^T (k+1)^{1/4}\bigg(\sum_{t=0}^{k-1}\a_t\bigg)^2 \leq \frac{64}{5\mu^2}(\ln T)^2 (T+1)^{5/4},
\end{equation}
Using the inequalities above, we get:
\begin{multline}\label{eq:gamma_term}
\sum_{k=1}^T(k+1)^2\Gamma_k \leq \frac{16}{\mu}T + \Big[ \frac{(128/3\mu^2)}{(1-\sigma_2)}(\ln T)^2 (T+1)^{3/2} \\
+\frac{\big(256n^2(L+8L^2/\mu)/\mu\big)}{(1-\sigma_2)^2}(\ln T)^2 (T+1)^{5/4} \Big] \Big(\frac{Cd}{2^b-1}\Big)^2\\ +
\frac{\big(8L(L+8L^2/\mu)\big)}{3\mu^3}(T+1)^{3/2}.
\end{multline}

Using \cref{eq:gamma_term} in \cref{eq:lyapunov_ineq2} and dividing both sides of the resulting inequality by $\sum_{k=1}^T(k+1)\approx (T+1)^2$, we get:

\begin{align*}
\mathbf{E}\big[\mathcal{V}_{T+1} \big] \leq  \frac{\mathbf{E}\big[\mathcal{V}_1\big]}{(T+1)^2} - 
  \frac{\frac{8}{\mu}\sum_{k=1}^T(k+1) \big( f(x^l_k)-f(x^*) \big)}{\sum_{k=1}^T(k+1)}\\
    + \frac{16}{\mu}\frac{T}{(T+1)^2} + \Big[\frac{(128/3\mu^2)}{(1-\sigma_2)}\frac{(\ln T)^2}{(T+1)^{1/2}}  \\
+\frac{\big(256n^2(L+8L^2/\mu)/\mu\big)}{(1-\sigma_2)^2}\frac{(\ln T)^2}{(T+1)^{3/4}} \Big]\Big(\frac{Cd}{2^b-1}\Big)^2 \\ +
\frac{\big(8L(L+8L^2/\mu)\big)}{3\mu^3}\frac{1}{(T+1)^{1/2}},
\end{align*}
which by rearranging both sides, dropping the positive term $\mathbf{E}\big[\mathcal{V}_{T+1} \big]$, and using Jensen's inequality, we obtain 
\begin{align}
&\mathbf{E}[f(z_{i}^{k})] - f^* \leq  \frac{\sum_{k=1}^T(k+1) \big( f(x^l_k)-f(x^*) \big)}{\sum_{k=1}^T(k+1)}\notag\\
&\leq \frac{\mu\mathbf{E}\big[\mathcal{V}_1\big]}{8(T+1)^2}
    + \frac{2}{T+1} + \frac{16 }{3\mu(1-\sigma_2)}\Big(\frac{Cd}{2^b-1}\Big)^2\frac{(\ln T)^2}{(T+1)^{1/2}}\notag  \\
&\quad +\frac{4n^2(L+8L^2)}{(1-\sigma_2)^2}\Big(\frac{Cd}{2^b-1}\Big)^2\frac{(\ln T)^2}{(T+1)^{3/4}}\notag  \\ 
&\quad +
\frac{\big(8L(L+8L^2/\mu)\big)}{3\mu^3}\frac{1}{(T+1)^{1/2}},\label{eq:rate}
\end{align}
where $z_{i}^{k}$ is defined in Algorithm \eqref{alg:QDSG}. Clearly, the right-hand side of \cref{eq:rate} is dominated by a term that decays with 
\begin{equation*}
    \mathcal{O}\bigg(\frac{(\ln T)^2}{(1-\sigma_2)(T+1)^{1/2}} \bigg).
\end{equation*}
\end{proof}

\section{A numerical example}


In this section, we provide a numerical experiment to illustrate for the performance of  \cref{eq:algorithm} under the random quantization scheme given in \cref{sec:increasing_quantizer}. Our simulation is similar to the ones in \cite{Doan:2018a}, however, we do not require any projection to a predefined compact set. As a result, the quantization intervals at each step in our algorithm are different and increasing in $k$.

We apply \cref{eq:algorithm} for solving the distributed variant of the popular linear regression problem. The goal of this problem is to find a linear relationship between a set of variables and some real valued outcome. That is, given a training set $S = \{(w_i,b_i)\in\mathbb{R}^{d}\times\mathbb{R}\}$ for $i=1,\cdots,n$. We want to find the solution of the following optimization problem
\begin{align}
\min_{x\in\mathbb{R}^{d}}\sum_{i=1}^n (w_i^Tx-b_i)^2,
\end{align}
In this case, we know that the optimal centralized solution can be efficiently found by solving a least squares problem. The challenge is to how to solve it efficiently in a decentralized fashion, under strict quantization constraints.

Suppose that $w_i\in\mathbb{R}^{5}$. We consider simulated training data sets where $(w_i,b_i)$, $i\in\{1,\cdots,n\}$, are generated randomly and independently according to the following uniform distributions:
\begin{equation}
    w_i(j) \sim \mathcal{U}[0,0.65],  \ \ j\in\{1,\cdots,d\}
\end{equation}
and
\begin{equation}
    b_i \sim \mathcal{U}[0,0.45].
\end{equation}

\begin{figure}[t!]
    \centering
    \includegraphics[width=0.8\columnwidth]{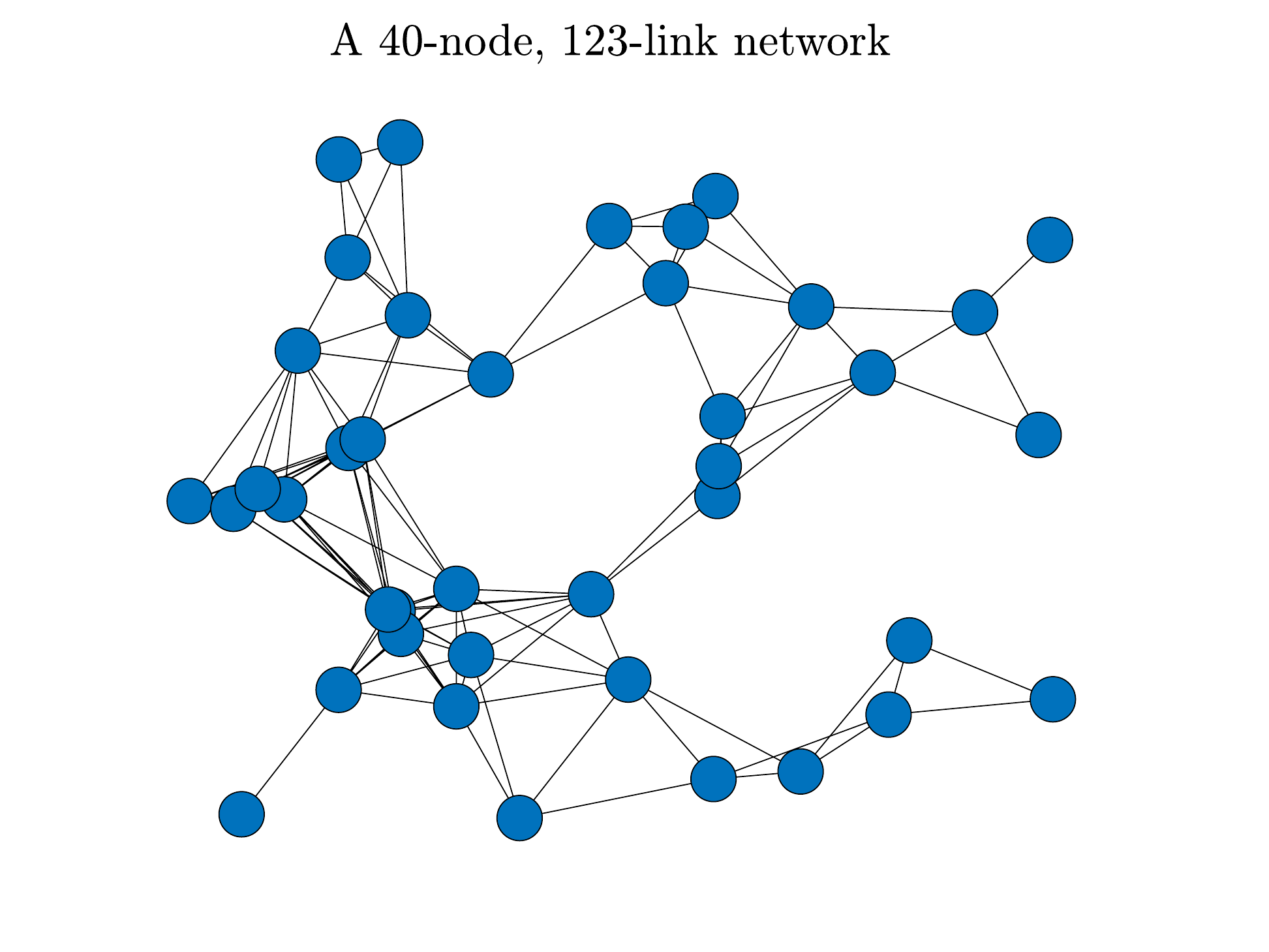}
    \caption{Network with 40 nodes and 123 communication links used in our distributed linear regression application.}
    \label{fig:graph_example}
\end{figure}

We study the performance of our distributed gradient method on the undirected connected graph of $40$ nodes shown in \cref{fig:graph_example}, i.e., $\mathcal{G}=(\mathcal{V},\mathcal{E})$ and $n = |\mathcal{V}|=40$. Our graph is generated randomly, similar to the ones in \cite{Doan:2018a}. Finally, the adjacency matrix $A$ is chosen as a Lazy Metropolis matrix corresponding to $\mathcal{G}$, i.e.,
\begin{align}
\mathbf{A} = [a_{ij}] = \left\{\begin{array}{ll}
\frac{1}{2(\max\{|\mathcal{N}_i| , |\mathcal{N}_j|\})}, & \text{ if } (i,j) \in \mathcal{E}\\
0, &\text{ if } (i,j)\notin\mathcal{E} \text{ and } i\neq j\\
1-\sum_{j\in\mathcal{N}_i}a_{ij},& \text{ if } i = j
\end{array}\right.\nonumber
\end{align}

\begin{figure}[t!]
    \centering
    \includegraphics[width=\columnwidth]{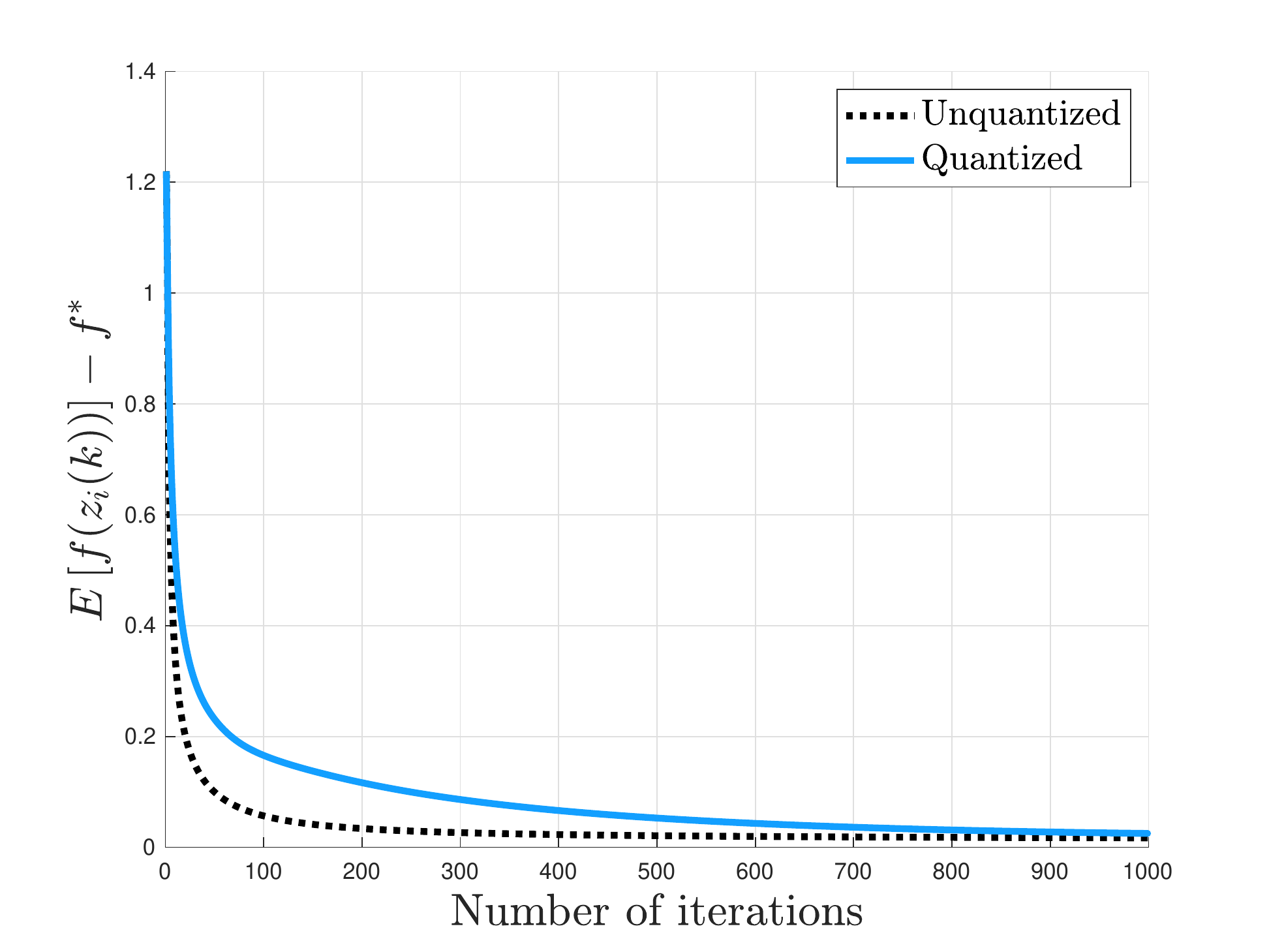}
    \caption{Performance of our algorithm with (solid line) and without (dashed line) quantization. In this example, the state variables was quantized using 16 bits/dimension.}
    \label{fig:my_label}
\end{figure}

In this example, the state variables $x^i_k$ were quantized using $16$ bits, without the need for periodic transmissions real numbers among agents. In \cref{fig:my_label}, we clearly see the convergence of the quantized vs. the unquantized algorithm, which asymptotically achieve the same value of the objective function, but there is a slight degradation due to the quantization error. It is important to note that despite the fact that the instantaneous quantization error increases with time, our algorithm successfully compensates for that by driving the estimates swiftly to the desired optimal solution. Hence, the excellent asymptotic convergence.


\section{Conclusions}
We introduced a novel two-time scale algorithm for distributed gradient descent under random quantization. Unlike other works, our algorithm does not require periodic transmission of real valued messages that specify the quantization bins for encoding and decoding, which in principle violates the finite bandwidth constraint at each link. We showed that the convergence rate of our algorithm is $\mathcal{O}((\ln(k))^2/\sqrt{k}),$ which improves on the current state of the art without the use of adaptive quantization.


\appendices

\section{Useful inequalities}
\begin{proposition}\label{prop:very_useful}
Let $a,b\in\mathbb{R}^d$. For $\eta>0$, the following inequality holds:
\begin{equation}\label{eq:CS1}
    \|a+b\|^2 \leq (1+\eta)\|a\|^2+(1+\frac{1}{\eta})\|b\|^2.
\end{equation}
or, equivalently,
\begin{equation}
\label{eq:CS2}
    \langle a,b\rangle \leq \frac{1}{2}\Big(\eta \|a\|^2+ \frac{1}{\eta} \|b\|^2\Big).
\end{equation}
\end{proposition}
\vspace{5pt}

\section{Proof of Lemma 3}
Using Lemmas 2 and 3, after some algebraic manipulations, we obtain
\begin{multline}\label{eq:aux}
    \mathbf{E}_{\mathcal{F}_k}[\mathcal{V}_{k+1}] \leq \Big(1-\frac{\mu}{2}\alpha_k\Big) \mathcal{V}_k + 2\a_k\big(f^\star - f(x^ l_k)\big)\\
    + \Big[ \a_k^2L^2 + \big(\b^2_k + 2\eta(1-(1-\sigma_2)\b_0)\a_k\b_kn^2 \big)\Big(\frac{Cd}{2^b-1}\Big)^2 \Big(\sum_{t=0}^{k-1}\a_t \Big)^2 \\ + \eta\Big(\frac{(1-\sigma_2)\b_0+1}{1-\sigma_2} \Big)L^2 \frac{\a_k^3}{\b_k^2}\Big]\\
    +\|Y_k\|^2_F \Big(L+\frac{8L^2}{\mu}  + \frac{\mu}{2}\eta_k - \eta(1-\sigma_2) \Big)\a_k.    
\end{multline}
Choosing $\eta=2(L+L^2/8)/(1-\sigma_2)$, since $\a_0/\b_0 =\mu/(1-\sigma_2)$, we may drop the last term in \cref{eq:aux}. Thus,
\begin{multline}\label{eq:aux2}
    \mathbf{E}_{\mathcal{F}_k}[\mathcal{V}_{k+1}] \leq \Big(1-\frac{\mu}{2}\alpha_k\Big) \mathcal{V}_k + 2\a_k\big(f^\star - f(x^ l_k)\big)\\
    + \Big[ \a_k^2L^2 + \big(\b^2_k + 2\eta(1-(1-\sigma_2)\b_0)\a_k\b_kn^2 \big)\Big(\frac{Cd}{2^b-1}\Big)^2 \Big(\sum_{t=0}^{k-1}\a_t \Big)^2 \\ + \eta\Big(\frac{(1-\sigma_2)\b_0+1}{1-\sigma_2} \Big)L^2 \frac{\a_k^3}{\b_k^2}\Big]
\end{multline}
Substituting the step sequences $\a_k$ and $\b_k$ into \cref{eq:aux2}, after some algebra, and taking the expectation over all possible $\mathcal{F}_k$ on both sides, we obtain \cref{eq:lyapunov_ineq}.

\bibliographystyle{IEEEtran}
\bibliography{IEEEabrv,refs}

\begin{thebibliography}{10}
\providecommand{\url}[1]{#1}
\csname url@samestyle\endcsname
\providecommand{\newblock}{\relax}
\providecommand{\bibinfo}[2]{#2}
\providecommand{\BIBentrySTDinterwordspacing}{\spaceskip=0pt\relax}
\providecommand{\BIBentryALTinterwordstretchfactor}{4}
\providecommand{\BIBentryALTinterwordspacing}{\spaceskip=\fontdimen2\font plus
\BIBentryALTinterwordstretchfactor\fontdimen3\font minus
  \fontdimen4\font\relax}
\providecommand{\BIBforeignlanguage}[2]{{%
\expandafter\ifx\csname l@#1\endcsname\relax
\typeout{** WARNING: IEEEtran.bst: No hyphenation pattern has been}%
\typeout{** loaded for the language `#1'. Using the pattern for}%
\typeout{** the default language instead.}%
\else
\language=\csname l@#1\endcsname
\fi
#2}}
\providecommand{\BIBdecl}{\relax}
\BIBdecl

\bibitem{Nedic:2020}
A.~{Nedic}, ``Distributed gradient methods for convex machine learning problems
  in networks: Distributed optimization,'' \emph{IEEE Signal Processing
  Magazine}, vol.~37, no.~3, pp. 92--101, 2020.

\bibitem{Yang:2019}
T.~Yang, X.~Yi, J.~Wu, Y.~Yuan, D.~Wu, Z.~Meng, Y.~Hong, H.~Wang, Z.~Lin, and
  K.~H. Johansson, ``A survey of distributed optimization,'' \emph{Annual
  Reviews in Control}, vol.~47, pp. 278--305, 2019.

\bibitem{Li:2017}
J.~Li, G.~Chen, Z.~Wu, and X.~He, ``Distributed subgradient method for
  multi-agent optimization with quantized communication,'' \emph{Mathematical
  Methods in the Applied Sciences}, vol.~40, no.~4, pp. 1201--1213, 2017.

\bibitem{ReisizadehMHP2019}
A.~{Reisizadeh}, A.~{Mokhtari}, H.~{Hassani}, and R.~{Pedarsani}, ``An exact
  quantized decentralized gradient descent algorithm,'' \emph{IEEE Transactions
  on Signal Processing}, vol.~67, no.~19, pp. 4934--4947, 2019.

\bibitem{pmlr-v119-taheri20a}
H.~Taheri, A.~Mokhtari, H.~Hassani, and R.~Pedarsani, ``Quantized decentralized
  stochastic learning over directed graphs,'' in \emph{Proceedings of the 37th
  International Conference on Machine Learning}, vol. 119, 13--18 Jul 2020, pp.
  9324--9333.

\bibitem{9157925}
T.~T. {Doan}, S.~T. {Maguluri}, and J.~{Romberg}, ``Fast convergence rates of
  distributed subgradient methods with adaptive quantization,'' \emph{IEEE
  Transactions on Automatic Control}, pp. 1--1, 2020.

\bibitem{Doan:2018a}
------, ``Convergence rates of distributed gradient methods under random
  quantization: A stochastic approximation approach,'' \emph{IEEE Transactions
  on Automatic Control}, 2020.

\bibitem{CM20}
A.~{Chattopadhyay} and U.~{Mitra}, ``Dynamic sensor subset selection for
  centralized tracking of an iid process,'' \emph{IEEE Transactions on Signal
  Processing}, vol.~68, pp. 3209--3224, 2020.

\bibitem{Doan:2018b}
T.~T. {Doan}, S.~T. {Maguluri}, and J.~{Romberg}, ``On the convergence of
  distributed subgradient methods under quantization,'' in \emph{2018 56th
  Annual Allerton Conference on Communication, Control, and Computing
  (Allerton)}, 2018, pp. 567--574.

\bibitem{Cinlar:2011}
E.~Cinlar, \emph{Probability and stochastics}.\hskip 1em plus 0.5em minus
  0.4em\relax Springer Science \& Business Media, 2011, vol. 261.

\bibitem{nedic2018network}
A.~Nedi{\'c}, A.~Olshevsky, and M.~G. Rabbat, ``Network topology and
  communication-computation tradeoffs in decentralized optimization,''
  \emph{Proceedings of the IEEE}, vol. 106, no.~5, pp. 953--976, 2018.

\end{thebibliography}

\end{document}